\documentclass[twocolumn,showpacs]{revtex4}

\usepackage{graphicx}%
\usepackage{dcolumn}
\usepackage{amsmath}

\makeatletter
\def\btt#1{\texttt{\@backslashchar#1}}%
\DeclareRobustCommand\bblash{\btt{\@backslashchar}}%
\makeatother

\begin{document}

\preprint{MgB2 1/T1}

\title[Short Title]{Pressure-induced anomalous magnetism and unconventional superconductivity in CeRhIn$_5$:$^{115}$In-NQR study under 
pressure}% Force line breaks with \\

\author{T.~Mito$^1$, S.~Kawasaki$^1$, G.~-q.~Zheng$^1$, Y. Kawasaki$^1$, K. Ishida$^1$, Y.~Kitaoka$^1$, D. Aoki$^2$, Y. 
Haga$^3$,  and Y. Onuki$^2$}

\affiliation{
$^1$Department of Physical Science, Graduate School of Engineering Science, Osaka University, Toyonaka, Osaka 560-8531, Japan\\
$^2$Department of Physics, Graduate School of  Science, Osaka University, Toyonaka, Osaka 560-0043, Japan\\
$^3$Advanced Science Research Center, Japan Atomic Energy Research Institute, Tokai, Ibaraki 319-1195, Japan}

\date{March 12, 2001}% It is always \today, today, but you may specify any date with \date.

\begin{abstract}
We report $^{115}$In nuclear-quadrupole-resonance (NQR) measurements of the pressure($P$)-induced superconductor CeRhIn$_5$ in the antiferromagnetic (AF) and superconducting (SC) states.
In the AF region, the internal field $H_{int}$ at the In site is substantially reduced from $H_{int}=1.75$ kOe at $P=0$ to 0.39 kOe at $P=1.23$ GPa, while the N\'eel temperature slightly changes with increasing $P$.
This suggests that either the size in the ordered moment $M_{Q}(P)$ or the angle $\theta (P)$ between the direction of $M_{Q}(P)$  and the tetragonal $c$ axis is extrapolated to zero at $P^*=1.6 \pm 0.1$ GPa at which a bulk SC transition is no longer emergent.
In the SC state at $P=2.1$ GPa, the nuclear spin-lattice relaxation rate $^{115}(1/T_1)$  has revealed a $T^3$ dependence 
without the coherence peak just below $T_c$, giving evidence for the unconventional superconductivity.
The dimensionality of the magnetic flutuations in the normal state are also discussed.
 
\end{abstract}

% PACS, the Physics and Astronomy Classification Scheme.
%\keywords{Suggested keywords}
                       
\maketitle
There is increasing evidence that a superconducting (SC) order in cerium (Ce)-based heavy-fermion (HF) compounds takes place 
nearby the border where an antiferromagnetic (AF) order is suppressed by applying pressure ($P$) to the HF-AF compounds 
CeCu$_2$Ge$_2$,\cite{CeCu2Ge2} CePd$_2$Si$_2$\cite{CePd2Si2} and CeIn$_3$.\cite{CeIn3}
When a magnetic medium is near an AF phase, AF waves of electron-spin density tend to propagate over a long distance with a low characteristic energy. Thereby it was argued that the binding of the Cooper pairs could be described in terms of the emission and absorption of fluctuating AF waves.\cite{CeIn3} The interplay between the AF  and SC states in the Ce-based HF systems may share some common aspects with other strongly-correlated-electron systems, and the understanding of mechanisms for superconductivity different from the conventional electron-phonon mediated one is still an important and unresolved issue.

Quite recently, it was discovered that a new HF-AF CeRhIn$_5$ (the N\'eel temperature, $T_N$=3.8 K) becomes a bulk HF superconductor at pressures exceeding $P_c\sim$1.63 GPa.\cite{Heeger}
It was suggested that a first order-like transition from an AF state to a SC state occurs with a SC transition temperature $T_c\sim 2.2$ K that is nearly 10 times larger than the maximum values for CePd$_2$Si$_2$ and CeIn$_3$.
Apparently, the evolution from the AF to SC states differs from all previous examples. 
$T_N$ was reported to increase weakly with the pressure for $P\leq$ 1.45 GPa, 
above which there is no resistive signature for $T_N$.
In order to shed light on a $P$-induced exotic evolution from the AF to SC states in CeRhIn$_5$, one needs to uncover magnetic and SC characteristics through extensive experiments under $P$.

Here we report extensive $^{115}$In-NQR measurements of CeRhIn$_5$ in the AF and SC states. The temperature ($T$) and $P$ dependences of the $^{115}$In-NQR spectrum and the nuclear spin-lattice relaxation rate $^{115}(1/T_1)$ were measured in a $P$ range of $0-2.1$ GPa and a $T$ range of 0.15$-$50 K. The salient results are
(1) $T_N$ slightly increases up to $P=1.00$GPa, but decreases at $P$=1.23 GPa,
(2) By contrast, the internal field $H_{int}$ at the In site due to the magnetic ordering is substantially reduced with $P$. This $P$-induced reduction in $H_{int}$ might be attributed to either an ordered moment $M_{Q}(P)\rightarrow 0$ or the angle $\theta\rightarrow 0$ at $P^*= 1.6\pm 0.1$GPa, where $\theta$ is the angle between the direction of $M_{Q}(P)$ and the tetragonal $c$ axis,
(3) The $^{115}(1/T_1)$ in the SC state at $P=2.1$ GPa obeys a $T^3$ dependence without the coherence peak just below $T_c$\, consistent with a line-node gap model as reported in all previous HF-SC compounds, \cite{HFSC}
(4) In the normal state at $P$=2.1GPa, the $T$ dependence of $^{115}(1/T_1)$ is consistent with the three dimensional (3D) SCR theory for a nearly AF Fermi-liquid state.

Single crystal of CeRhIn$_5$ was grown by the  self-flux method.\cite{Heeger} Powder X-ray diffraction indicated that the 
compound consists of a single phase that is formed in the primitive tetragonal HoCoGa$_5$ structure. The single crystal was 
moderately crushed into grains in order to make rf pulses penetrate into samples easily. Hydrostatic pressure was applied by 
utilizing a NiCrAl/BeCu piston-cylinder cell, filled with a Si-based organic liquid as a pressure-transmitting medium. 
High-frequency ac-susceptibility (ac-$\chi$) was measured at $P$=1.5, 1.65, 1.8 and 2.15 GPa by using the {\it in-situ} NQR coil.
The ac-$\chi$ data show a sharp SC transition at $T_c$=2.1 and 2.2 K at $P$=1.8 and 2.15 GPa, respectively (see Fig.4(b)).
The $^{115}$In-NQR spectrum was obtained by plotting spin-echo signal as a function of frequency in  $T=1.4 -10$ K and $P=0-2.1$ 
GPa. $T_1$ was measured by the conventional saturation-recovery method in $T=0.15-50$ K at $P$=0, 1.23 and 2.1 GPa.

CeRhIn$_5$ consists of alternating layers of CeIn$_3$ and RhIn$_2$ and hence has two inequivalent In sites per a unit cell. The 
In(1) site, analogous to the single In site in a cubic CeIn$_3$, is located on the top and bottom faces of the tetragonal unit cell. By considering the symmetry of the In(1) site, this site was characterized  by  $\nu_Q=6.78\pm 0.01$ MHz and an asymmetry 
parameter $\eta$=0 in the previous NQR study.\cite{Curro}  Here $\nu_{Q}$ and $\eta$ are defined by the NQR Hamiltonian: 
$H_Q=(h\nu_Q/6)[3I_z^2-I^2+\eta(I_x^2-I_y^2)]$. The In(1)-NQR spectrum in the paramagnetic state at 4.2 K and $P=0$ is shown in the bottom of Fig.1 where four transitions are found at the different frequencies $\nu=n\nu_Q$ with $n$=1, 2, 3 and 4, respectively.

In order to deduce the $T$ dependence of the internal field $H_{int}(T)$ at the In(1) site in the AF state, we focus on the splitting of the 2$\nu_Q$ (3/2$\Longleftrightarrow$1/2) transition and the shift of resonance frequency of the 3$\nu_Q$ (5/2$\Longleftrightarrow$3/2) transition in the In(1) spectrum below $T_N$.
This is because $H_{int}$ lies perpendicular to the tetragonal $c$ axis as reported in the previous study.\cite{Curro}
Including the Zeeman term $H_Z=-\gamma\hbar I_x\cdot H_{int}$ where $\gamma$ is the gyromagnetic ratio, we diagonalize the full Hamiltonian $H_{nuc}=H_Q+H_Z$ and determine the $H_{int}(T) $ for different values of $P$.
The 2$\nu_Q$ transition at $P=0$ shown in the upper panel of Fig.1 is asymmetrically split into two resonances by $H_{int}$, consistently with the previous result.\cite{Curro}
By contrast, the resonance frequency $\nu_p$ of the 3$\nu_Q$ transition is decreased by $H_{int}$ as seen in Fig.1.
The $T$ dependence of $\nu_p$ at $P$=0, 0.46, 1.00 and 1.23 GPa is shown in Fig.2(a).
$T_N$ is marked by arrows in Fig.2(a) and is precisely determined as the temperature below which $\nu_p$ decreases.
It is notable that $T_N$ slightly increases from 3.8 K at $P$=0 to $\sim$ 4 K at $P=1.00$ GPa, but decreases to $\sim$ 3.6 K at $P$=1.23 GPa.
The occurrence of the magnetic ordering at $P=1.23$ GPa is clearly corroborated by a distinct peak in $1/T_1T$ at $T_N=3.6$ K that probes critical magnetic fluctuations toward the magnetic ordering as shown in the inset of Fig.2(a).

$H_{int}=1.75$ kOe at $P=0$ and $T=1.4$ K is estimated from the size of $\nu_p$ reduction of the 3$\nu_Q$ transition as well as the splitting of the 2$\nu_Q$ transition. This is consistent with the previous result.\cite{Curro}
Note that $\nu_p$ is only sensitive to the magnitude of $H_{int}$. $H_{int}(P)$ is plotted against a reduced temperature $t=T/T_N$ in Fig.2(b).
Unexpectedly, the saturated value of $H_{int}\sim$ 0.39 kOe at $P=1.23$ GPa is about five times smaller than $H_{int}\sim$ 1.75 kOe at $P=0$, although $T_N$ changes moderately.
This slight pressure dependence of $T_N$ contrasts with the strong reduction of $H_{int}$.
A recent neutron experiment reported that Ce-ordered moments ($M_Q=0.264(4)\mu_B$ at 1.4 K and $P=0$) that lie in the basal plane are antiferromagnetically aligned, but they spiral transversely along the $c$ axis with an incommensurate wave vector {\bf q$_{\rm M}$}=(1/2, 1/2, 0.297).\cite{Bao}
$H_{int}(P)$ is then extrapolated to zero at $P^*=1.6 \pm 0.1$ GPa (see Fig.4(a)).
If $M_Q(P)$ is directed in the basal plane, the $M_Q$ would be scaled to $H_{int}$ and substantially reduced to $\sim 0.05\mu_B$ at $P=1.23$GPa.
On the other hand, if $M_Q(P)$ is rotated with $P$ from the $ab$ plane to the $c$ axis, the angle $\theta$ between the direction of $M_Q$ and the $c$ axis would be progressively smaller, extrapolated to zero at $P^*=1.6 \pm 0.1$ GPa (see Fig.4(a)).
This is because $H_{int}$ at the In(1) site is canceled out at $\theta$=0.
$H_{int}$ orignates from the direct dipolar field from the Ce ordered moments 
that reaches 30\% of the total and from the indirect "pseudo" dipolar (anisotropic) field via the supertransferred hyperfine interaction. 
The latter internal field acts on the In site through the hybridization between In 5$p$- and Ce 4$f$-orbits. Note that the isotropic hyperfine field originating from Ce ordered moments is canceled out at the In(1) site.
In order to see which $P$-induced change is more likely in the AF state, we need to consider the $T$ dependence of $H_{int}(t)/H_{int}(0)$ displayed in the inset of Fig.2(b), where $H_{int}(0)$ is the low-$T$ saturated value. A rapid growing of $H_{int}(T)$ is evident even at $P$=1.23 GPa.
It would be therefore unlikely that some itinerant magnetic ordering takes place with a {\it reduced moment} and rather likely that the ordered moments rotate toward the $c$ axis.
As a result, it might be expected that the spiral order evolves into some commensurate AF fluctuation regime at $P^*=1.6 \pm 0.1$ GPa. Note that $P^*$ is close in value to a critical pressure $P_c\sim 1.63$ GPa which was suggested from the resistivity measurement.\cite{Heeger}
To resolve this issue, further neutron experiment under pressure is highly desired.

We next deal with the SC region.
The $T_1$ in the SC and normal state at $P=2.1$ GPa was measured at the 1$\nu_Q$ and 2$\nu_Q$ transitions in order to avoid  heating effect due to rf-excitation pulses.
$T_1$ was determined by a single component.
Figure 3 shows the $T$ dependence of $^{115}(1/T_1)$ at $P=0$ and 2.1 GPa.
$^{115}(1/T_1)$ exhibits no coherence peak just below $T_c$=2.2 K, followed by a $T^3$ dependence down to $\sim$ 0.3 K.
This is a convincing experimental evidence for the unconventional nature of the $P$-induced superconductivity in CeRhIn$_5$.
Likewise all previous examples, a line-node gap model is applicable to the SC state in CeRhIn$_5$. Assuming an  anisotropic energy gap model with $\Delta=\Delta_0\cos{\theta}$, a solid line in Fig.3 is a fit for the $^{115}(1/T_1)$ data with $2\Delta_0=5 k_{\rm B}T_c$.

We argue magnetic characters in the normal state at $P=$2.1 GPa.
According to the SCR theory for nearly-AF metals by Moriya {\it et al},\cite{moriya} $1/T_1T\propto \chi_Q(T)^n$.
Here a power-law dependence  of the staggered susceptibility $\chi_Q(T)$ is obtained as $n$=1 and 1/2 for two (2D) and three (3D) dimensional electronic systems, respectively.
By noting that $\chi_Q(T)$ follows a Curie-Weiss law of $1/(T+\theta)$, a behavior of $(T_1T)^2\propto (T+\theta)$ is expected for the 3D nearly AF regime.
As a matter of fact, as indicated in the inset of Fig.3, a fit of $(T_1T)^2\propto (T+\theta)$ with $\theta=1.5$ K is consistent with the present result in a relatively wide $T$ range of $T_c=2.2$ K -- 30 K.
This shows that 3D AF fluctuations are dominant in the normal state at $P$=2.1GPa.\cite{moriya,ishigaki,nakamura}. 

Hegger {\it et al.} speculated that the maximum at $T_{\chi m}=$7.5 K and $P=0$ in the susceptibility is associated with the 
development of 2D AF correlations in the CeIn$_3$ layers,\cite{Heeger} since a 2D-like magnetic character is expected from  its 
quasi-2D crystal structure in the lower $P$ region. With increasing $P$, $T_{\chi m}$ decreases approximately linearly and it would be extrapolated to $T=0$ at $P_m=1.3\pm 0.4$ GPa. 
This is indicative of the 2D character in magnetic properties, which may be progressively suppressed as $P$ increases. It is 
noteworthy that $P_m$ is comparable to $P_c\sim 1.63$ GPa and $P^*= 1.6 \pm 0.1$ GPa.
Therefore, at $P$ = 2.1GPa exceeding either $P_m$ or $P^*$, it is considered  that AF fluctuations possess a 3D character rather than a 2D one. However, further works are needed to elucidate the role of critical AF fluctuations in the onset of the unconventional $P$-induced superconductivity in CeRhIn$_5$.

Figure 4(a) presents a phase diagram of the AF and SC phases along with  the previous results.\cite{Heeger}
According to the Ref.[4], at $P_c\sim$1.63 GPa, the SC transition in the resistivity measurement begins around 2 K and reaches a zero-resistance state with a broad transition width.\cite{Heeger}
In agreement, as shown in Fig.4(b), we found that the onset temperature in ac-$\chi$ is in accord with this zero-resistance $T_c$.
However the size of the SC diamagnetism at 1.4 K in ac-$\chi$ is substantially reduced.
At $P=1.5$ GPa, no change in ac-$\chi$ is observable at all, supporting a critical pressure $P_c\sim$1.63 GPa as suggested in the previous work.\cite{Heeger}
Therefore it was ensured from the present ac-$\chi$ measurement that the bulk SC transition takes place down to $P=1.8$ GPa, 
but probably not at pressures lower than $P=1.65$ GPa.
The present NQR study confirms that $T_N$ slightly increases up to $T_N\sim$ 4 K at $P$=1.00 GPa, but decreases to 3.6 K at $P$=1.23 GPa.
The internal field $H_{int}$ is extrapolated to zero at $P^*=1.6 \pm 0.1$ GPa which is close to $P_c\sim$1.63 GPa. Either the reduction in the ordered moment $M_{Q}$ or its rotation from the $ab$ plane to the $c$ axis may occur as $P$ increases. From the rapid growing of $H_{int}(P)$ upon cooling below $T_N$, we believe that the rotation of $M_Q(P)$ occurs with $P$, but its marked reduction does not.
In this context, we suggest that the spiral order is presumably suppressed across a critical pressure $P^*= 1.6\pm 0.1$ GPa.
Eventually, the SC transition emerges in CeRhIn$_5$ at pressures exceeding $P^*$. It is highly desired to elucidate whether AF or SC fluctuations prevent the onset of any type of long-range orders in the vicinity of $P^*=1.6\pm 0.1$ 
GPa. 

In conclusion, we have reported that unconventional magnetic and superconducting states are induced by applying $P$ to the HF-AF CeRhIn$_5$. In the magnetic region, $T_N$ exhibits a moderate variation. By contrast, $H_{int}$ whose presence is due to the magnetic ordering is unexpectedly reduced at $P=1.23$ GPa, extrapolated to zero at $P^*=1.6\pm 0.1$GPa. The spiral order might be suppressed presumably due to the rotation of the ordered moments toward the $c$ axis. This $P^*$ is comparable to $P_c\sim$1.63 GPa at which the bulk SC transition is not emergent as suggested from the previous resistivity \cite{Heeger} and corroborated by the present ac-$\chi$ measurements. In the SC state at $P=2.1$GPa, we found $1/T_1\propto T^3$ that shows the existence of line-nodes in the gap function. In the normal state, the remarkable behavior of $(1/T_1T)\propto1/\sqrt{T+1.5}$ which is consistent with the 3D nearly AF fluctuation regime suggests that the magnetic nature possesses a 3D-like character at pressures where the bulk SC sets in.
This work was supported by the COE Research (10CE2004) in Grant-in-Aid for Scientific Research from the Ministry of Education, 
Sport, Science and Culture of Japan. One of the authors (T.M.) has been supported by JSPS Research Fellowships for Young 
Scientists.

\begin{figure}[htbp]

\caption[]{$^{115}$In-NQR spectra at various values of pressure. The upper panel indicates the 2$\nu_Q$ and 3$\nu_Q$ transitions below $T_N$.

The lower panel indicates the $^{115}$In-NQR spectrum at ambient pressure 
($P=0$) above $T_N$.
In the absence of internal field, the In(1) spectrum consists of four transitions given by $\nu=n\nu_Q$, where $n=$1, 2, 3 
and 4 (see text).
Below $T_N$, the $2\nu_Q$ and 3$\nu_Q$ transition splits asymmetrically and shifts, respectively.}
\end{figure}

%fig2
\begin{figure}[htbp]

\caption[]{(a) Temperature dependence of the resonance frequency $\nu_p$ of the 3$\nu_Q$ transition at $P$=0, 0.46, 1.00 and 1.23 GPa.
The arrow indicates $T_N$. The inset shows the temperature dependence of $^{115}(1/T_1T)$ at $P$=1.23GPa.
(b) Temperature dependence of the internal field $H_{int}$ is plotted against $t=T/T_N$ at $P$=0, 0.46, 1.00 and 1.23 GPa.
The inset indicates  $H_{int}(t)/H_{int}(0)$ vs $t$ plots.
Here $H_{int}(0)$ is a saturated value at low temperature.}

\end{figure}

%fig3
\begin{figure}[htbp]

\caption[]{Temperature dependence of $^{115}$In nuclear spin-lattice relaxation rate, $^{115}(1/T_1)$ at $P=2.1$ GPa along with the data at $P=0$ both displayed in logarithmic scales.
The solid line is a fit assuming a line-node gap $\Delta(\phi)=\Delta_0\cos{\phi}$ with $2\Delta_0=5k_{\rm B}T_c$.
Inset: $(T_1T)^2$ vs $T$ plot at $P=2.1$ GPa.
The solid line is a fit based on the 3D-SCR theory \cite{moriya,ishigaki,nakamura} that predicts the following behavior:
$(T_1T)^2\propto 1/\chi_Q(T)\propto (T+\theta)$ where $\theta=1.5$ K.
Note that $\chi_Q\propto (T+\theta)^{-1}$ follows a Curie-Weiss law for the nearly AF Fermi-liquid regime.}

\end{figure}

%fig4
\begin{figure}[htbp]

\caption[]{(a) The pressure dependences of $T_N$ (open triangles), $T_c$ (open circles) and $H_{int}$ (open diamonds) determined from the present work are shown together with the previous data.\cite{Heeger}
The $H_{int}(P)$ is extrapolated to zero at $P^*=1.6\pm 0.1$GPa as indicated by the dotted line.
If the reduction of $H_{int}(P)$ is attributed to the rotation of $M_Q(P)$, $H_{int}(P)$ is proportional to sin$\theta$ (See text).
The SC transition width was marked by bars: $T_{\rm  c}^{\rm onset}-T_{\rm c}^{\rm offset}$ defined in Fig.4(b).
The solid lines are guides to the eye.

(b) Temperature dependence of the high-frequency ac susceptibility (ac-$\chi$) measured using an {\it in-situ} NQR coil at various values of pressure.
$T_{\rm c}$ is defined as the temperature at which the ac-$\chi$ decreases to 10\% of the total Meissner signal at each pressure.
$T_{\rm  c}^{\rm onset}$ and $T_{\rm c}^{\rm offset}$ are defined as the respective temperature at which the SC diamagnetism starts to emerge and the ac-$\chi$ reaches to 90\% of the total Meissner signal.}

\end{figure}

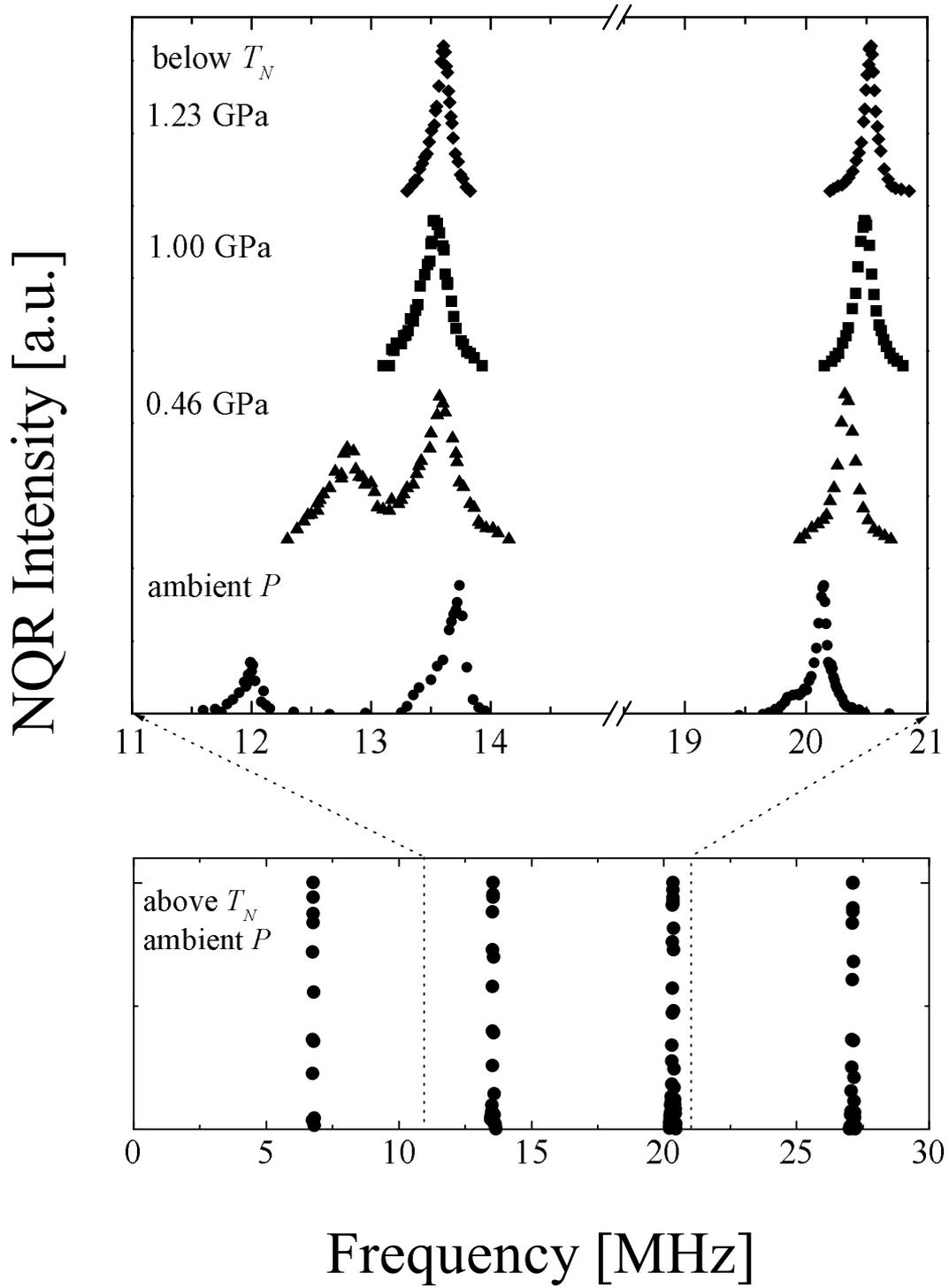
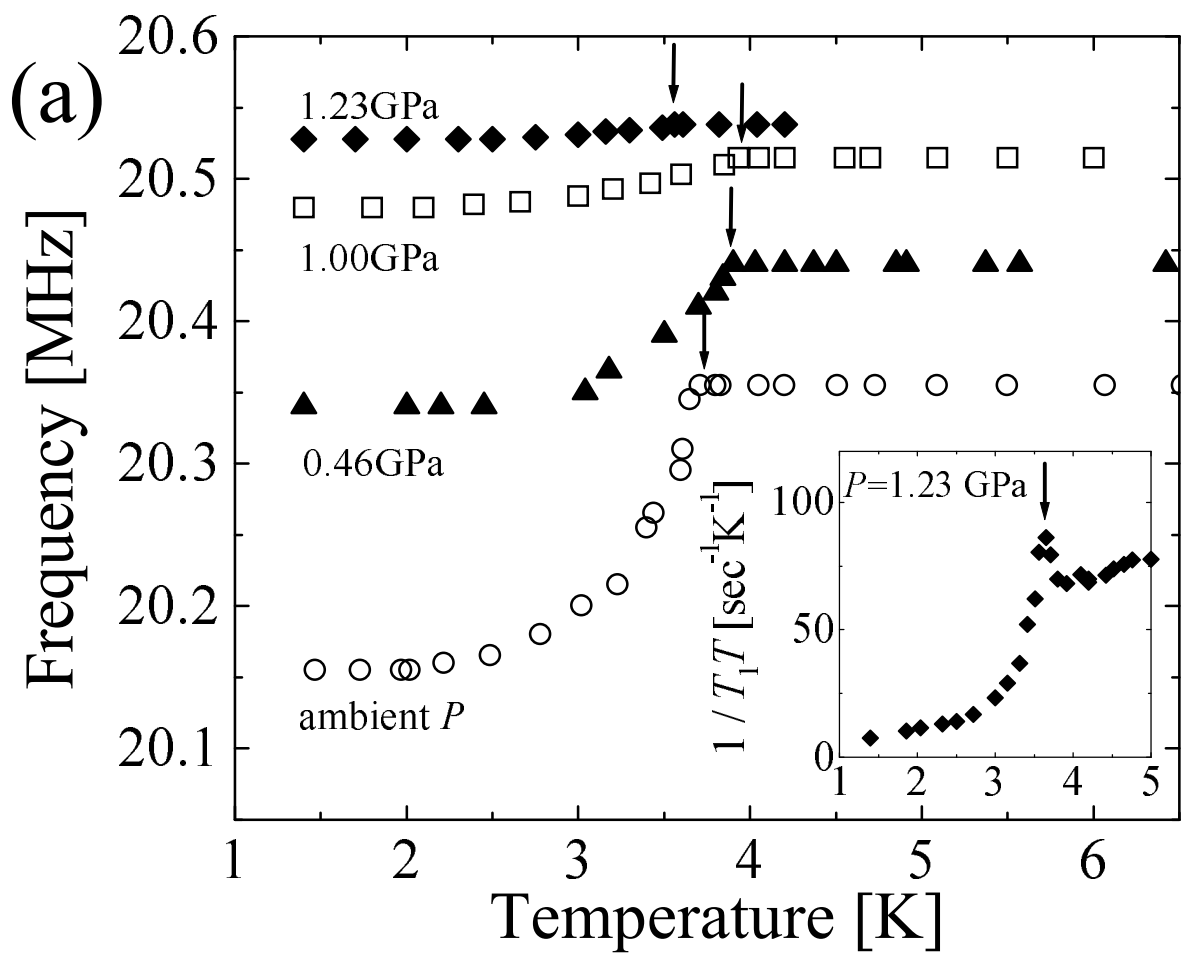
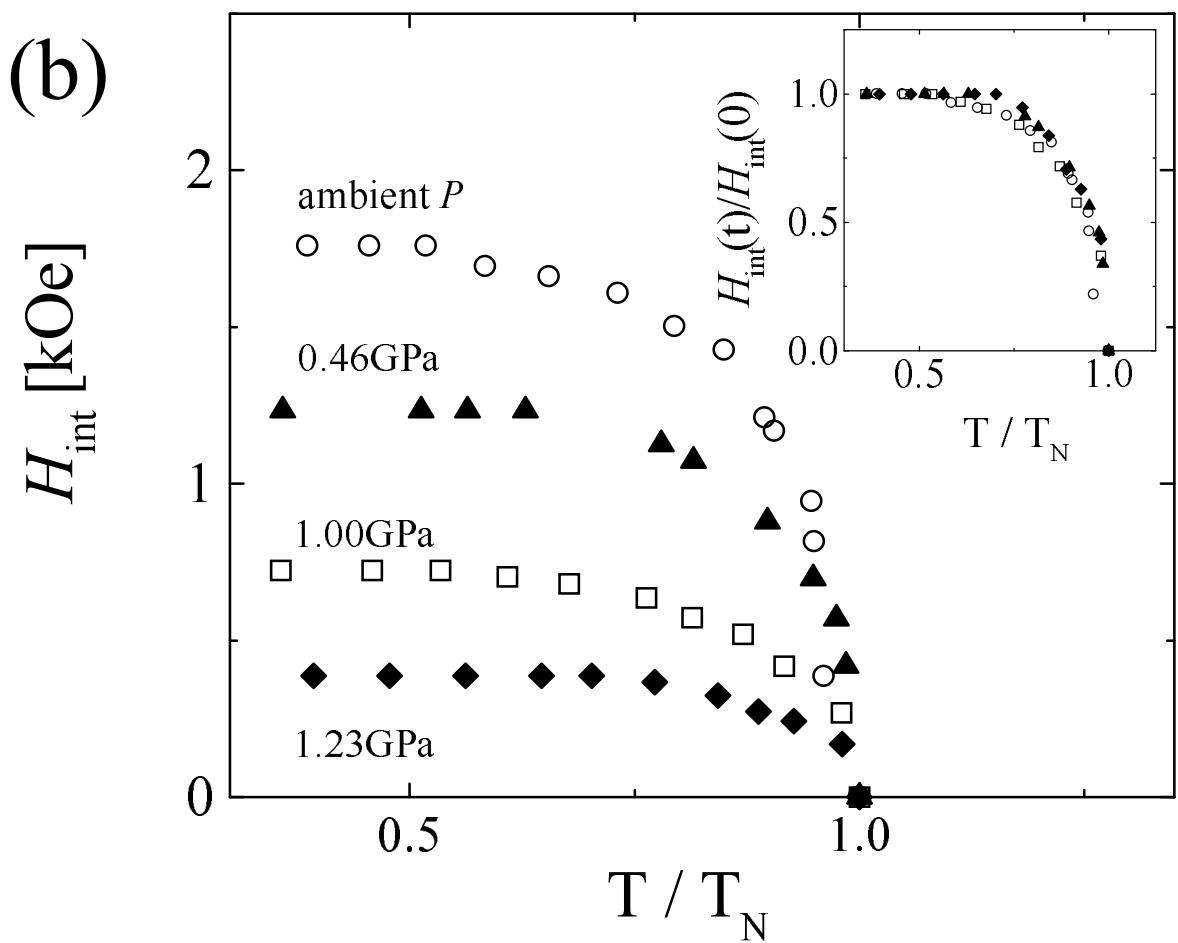
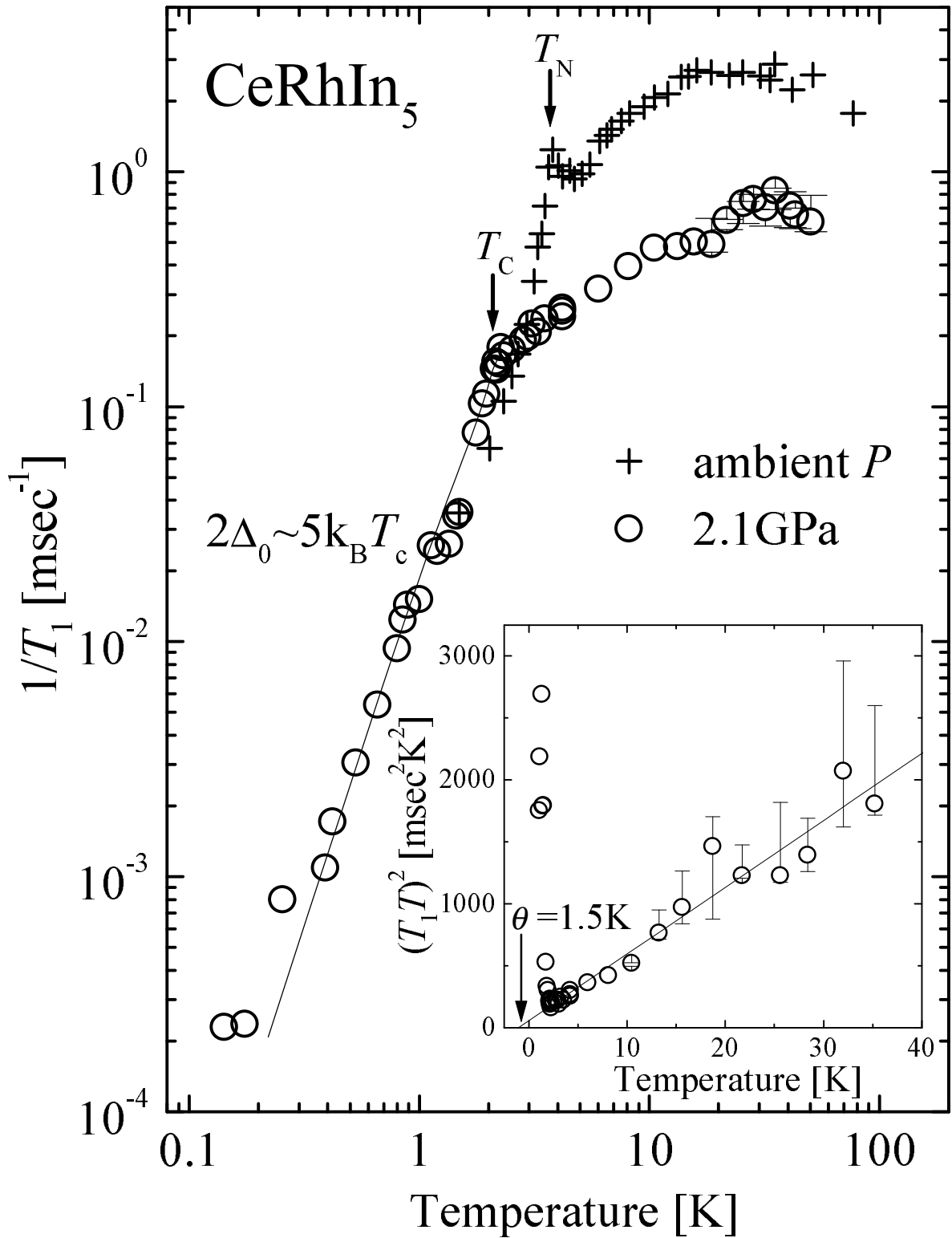
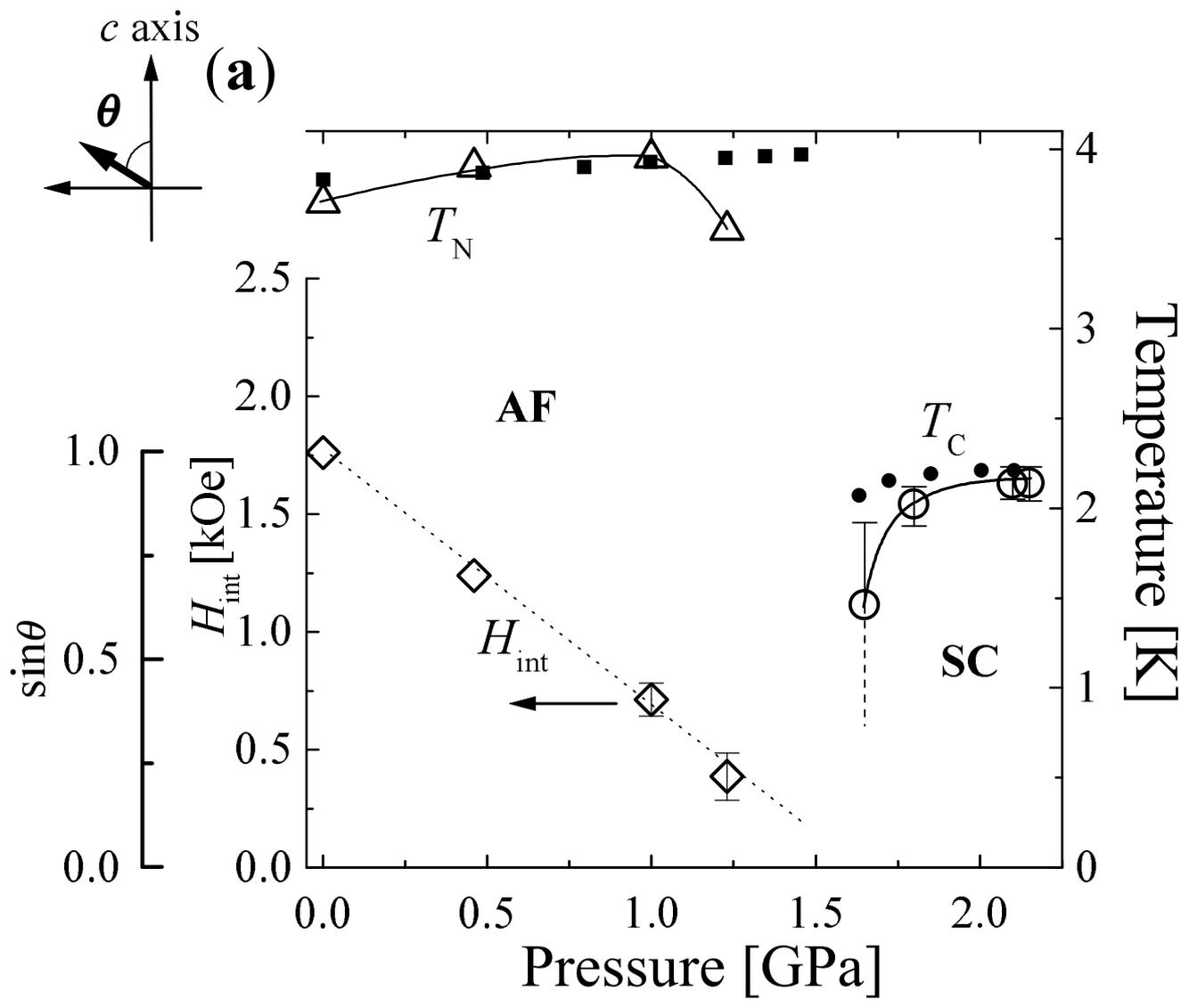
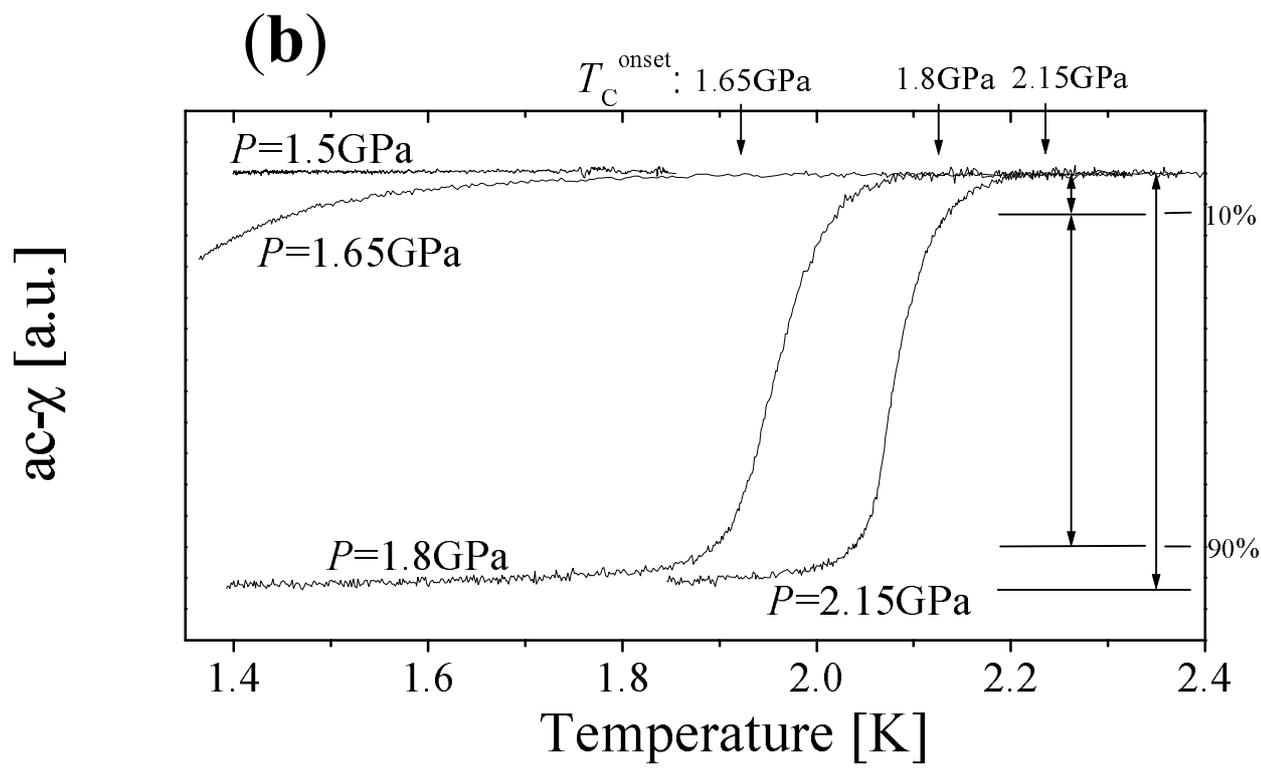
\end{document}